\documentclass[%
 reprint,
 amsmath,amssymb,
 aps,
]{revtex4-2}

\usepackage{graphicx}
\usepackage{dcolumn}
\usepackage{bm}

\usepackage{scalerel}
\usepackage{tikz}
\usetikzlibrary{svg.path}
\definecolor{orcidlogocol}{HTML}{A6CE39}
\tikzset{
  orcidlogo/.pic={
    \fill[orcidlogocol] svg{M256,128c0,70.7-57.3,128-128,128C57.3,256,0,198.7,0,128C0,57.3,57.3,0,128,0C198.7,0,256,57.3,256,128z};
    \fill[white] svg{M86.3,186.2H70.9V79.1h15.4v48.4V186.2z}
                 svg{M108.9,79.1h41.6c39.6,0,57,28.3,57,53.6c0,27.5-21.5,53.6-56.8,53.6h-41.8V79.1z M124.3,172.4h24.5c34.9,0,42.9-26.5,42.9-39.7c0-21.5-13.7-39.7-43.7-39.7h-23.7V172.4z}
                 svg{M88.7,56.8c0,5.5-4.5,10.1-10.1,10.1c-5.6,0-10.1-4.6-10.1-10.1c0-5.6,4.5-10.1,10.1-10.1C84.2,46.7,88.7,51.3,88.7,56.8z};
  }
}
\newcommand\orcidicon[1]{\href{https://orcid.org/#1}{\mbox{\scalerel*{
\begin{tikzpicture}[yscale=-1,transform shape]
\pic{orcidlogo};
\end{tikzpicture}
}{|}}}}
\usepackage{hyperref}

\begin{document}

\preprint{APS/123-QED}

\title{Constraining Electromagnetic Signals from Black Holes with Hair}

\author{Nicole R. Crumpler\orcidicon{0000-0002-8866-4797}}
\email{ncrumpl2@jh.edu}
\affiliation{William H. Miller III Department of Physics and Astronomy, Johns Hopkins University, Baltimore, MD 21210, USA}

\date{\today}

\begin{abstract}\label{sec:abstract}

\indent We constrain a broad class of ``hairy" black hole models capable of directly sourcing electromagnetic radiation during a binary black hole merger. This signal is generic and model-independent since it is characterized by the black hole mass ($M$) and the fraction of that mass released as radiation ($\epsilon$). For field energy densities surpassing the Schwinger limit, this mechanism triggers pair-production to produce a gamma-ray burst. By cross-referencing gravitational wave events with gamma-ray observations, we place upper bounds of $\epsilon<10^{-5}-10^{-4}$ for $10-50$ $M_\odot$ black holes depending on the black hole mass. We discuss the weak detection of a gamma-ray burst following GW150914 and show that this event is consistent with rapid electromagnetic emission directly from a ``hairy" black hole with $\epsilon\sim10^{-7}-10^{-6}$. Below the Schwinger limit, ambient charged particles are rapidly accelerated to nearly the speed of light by the strong electromagnetic field. For 1-50 $M_\odot$ black holes and $\epsilon$ ranging from $10^{-20}$ to $10^{-7}$, the typical proton energies are $\sim20$ GeV-20 TeV and electron energies are $\sim0.01-10$ GeV. At these energies, cosmic ray protons and electrons quickly diffuse into the Milky Way's background magnetic field, making it difficult to identify a point source producing them. Overall, constraining $\epsilon$ in this less energetic regime becomes difficult and future constraints may need to consider specific models of ``hairy" black holes.
\end{abstract}

\maketitle


\section{Introduction} \label{sec:intro}

\indent There is an important distinction between astrophysical and mathematical black holes (BHs). Astrophysically, BHs are observed as compact regions of spacetime in which gravity is so strong that even light cannot escape. These objects have been detected merging with each other \cite{Abbott_2019}, emitting electromagnetically-bright jets \cite{Vayner_2021}, consuming stars \cite{Komossa_2015}, and more. Mathematically, BHs are vacuum solutions of the Einstein equations of general relativity describing spacetime external to a compact mass distribution within an event horizon. The theory underlying mathematical BHs has been used to characterize astrophysical BHs, although there remain clear disconnects between these theoretical models and physical reality. 

\indent One such disconnect stems from the so-called ``no-hair" theorem. Canonically, the ``no-hair" theorem proposes that mathematical BHs are completely characterized by the BH's mass, charge, and spin as seen by an external observer. This theorem has been tested astrophysically using a variety of probes such as radio observations of the shadow of a BH event horizon \cite{Broderick_2014, Psaltis_2016, Wang_2022}, gravitational wave signals of binary BH (BBH) mergers \cite{Isi_2019, Wang_2022_gw}, and stellar orbits around the galactic center \cite{Sadeghian_2011, Qi_2021}. No evidence for its violation has yet been discovered in astrophysical BHs. However, the ``no-hair" theorem leads directly to the BH information paradox \cite{Hawking_1976}. In this paradox, different configurations of matter, radiation, etc. that have fallen into a BH can be described by the same mathematical BH solution, losing information about the initial quantum state of the system. This violates a core tenet of quantum mechanics in a regime in which both the theories of general relativity and quantum mechanics are valid. There have been many attempts to resolve this paradox; one can refer to Ref. \cite{Raju_2022} for a recent review. Despite these attempts, no consensus has yet been reached.

\indent The BH information paradox supports a possibility of richer physics underlying BHs. ``Hairy" BHs are novel solutions to the Einstein field equations which are characterized by more than the three parameters of a canonical BH. Some of these BH models have been proposed as explicit solutions to the BH information paradox. One interesting possibility, proposed by Ref. \cite{Kaplan_2019}, is the firewall BH model. This mathematical BH has a singular shell (otherwise known as a firewall) just outside the horizon, causing general relativity to break down outside the BH horizon. Since general relativity no longer holds in this regime, the BH information problem no longer applies. Such an exotic object appears as a Schwarzschild BH to a distant observer, raising the question of how a firewall BH (and, more generally, other ``hairy" BHs) might be distinguished from a canonical BH in astrophysical observations. 


\indent Electromagnetic (EM) radiation from astrophysical BHs in baryon-poor environments would be a beacon of new fundamental physics and support the existence of non-canonical BHs. Canonical BHs do not directly source EM radiation, except the weak emission of thermal Hawking radiation \cite{Hawking_1975}, which is not observable for BHs of astrophysically relevant masses. However, there are convincing qualitative arguments indicating that ``hairy" BH models could radiate an appreciable proportion of their mass as EM radiation. As a motivating example, consider the firewall BH discussed earlier. Ref. \cite{Kaplan_2019} suggests several ways in which this model could produce EM radiation including the explosion of an unstable firewall, a BH phase transition from a canonical to a firewall BH, and BBH mergers involving a firewall BH. We emphasize that no quantitative model for this effect yet exists, since working out the details of such a model would likely be a challenging exercise in quantum gravity. We hope this paper will convince those working in the field that such a pursuit is worthwhile, given the fact that these models generically produce observable signals and thus can be constrained via astrophysical observations. Overall, our understanding of BHs is inconsistent, motivating searches for generic signals of deviations from canonical BH models.

\indent Given that only non-canonical BHs can radiate appreciably, there are two important considerations in order to distinguish this radiation from typical astrophysical sources. Firstly, we must be confident that there is a BH in the region sourcing the radiation. Secondly, the BH must be in a sufficiently baryon-poor environment and the emitted radiation must be sufficiently energetic that we can be confident the radiation is not produced by standard processes such as relativistic jets. The best observable available satisfying these considerations is concurrent observations of BBH mergers with gravitational wave detectors and EM radiation instruments. Gravitational wave detectors such as the Advanced Laser Interferometer Gravitational-Wave Observatory (LIGO, \cite{LIGO_2015}) and Virgo \cite{Acernese_2015} regularly observe the mergers of stellar mass BHs. These events have only been observed extragalactically, with local BBH merger rates measured to be $\sim 10 \text{ Gpc}^{-3}\text{ yr}^{-1}$ \cite{Mandel_2016}. Thus, any observable EM radiation from such events must be extremely energetic, on the order of supernova energies. 

\indent There is typically insufficient baryonic matter surrounding BBHs to produce EM radiation observable at these extragalactic distances and energy scales. Some models have been proposed to produce a gamma-ray burst (GRB) during a BBH merger. Most of these models require rapid accretion during the merger \cite{Loeb_2016,Perna_2016,Woosley_2016}, necessitating a baryon-rich environment, and all of these models have been contested \cite{Dai_2017,Fedrow_2017, Kimura_2017}. Another class of model involves charged BHs \cite{Zhang_2016}, but the required charge is unreasonably large \cite{Lyutikov_2016}. Given the shortcomings of these models, no extragalactically-observable EM signal is expected from a stellar-mass BBH merger. Thus, in this paper, we investigate the observational signatures of a stellar-mass BH directly releasing some of its mass as EM radiation during a BBH merger as a novel indicator of the existence of ``hairy" BHs.

\indent Since BBHs mergers are catastrophic events characterized by short timescales, a large amount of energy could be emitted by the BH in a burst of EM radiation. We parameterize the amount of energy released by $\epsilon$ such that $\epsilon M$ is the energy emitted directly by the BH in EM radiation into the ambient environment. The characteristic frequency of the EM signal is independent of the ``hairy" BH model sourcing the radiation because any such emission directly from the BH must tunnel out of the gravitational well in the same manner as Hawking radiation \cite{Parikh_2000}. If the Schwarzschild radius of a BH is $r_s$, then the characteristic frequency of the radiation emitted by the BH is 
\begin{equation}
    f=\frac{1}{2 r_s}=3.3\times 10^{-17}\left(\frac{M}{M_\odot}\right)^{-1}\text{ MeV}.
\end{equation}
Throughout this paper, we work in natural units where $\hbar=c=k_B=\epsilon_0=1$ unless otherwise stated. The frequencies emitted by stellar mass BHs are very low-frequency radio waves. At these frequencies, all of the radiation is absorbed in the interstellar medium, predominantly via free-free absorption by the warm ionized medium \cite{Reynolds_1990}. Thus, this radiation is not directly observable, but is absorbed and re-emitted as a secondary signal. We calculate the range of $M$ and $\epsilon$ for which this secondary signal could be detected. In future work, model-dependent effects will need to be included to augment this generic parameterization. Again we emphasize that, although there are ``hairy" BH models capable of producing EM radiation, no complete model able to make quantitative predictions for this effect exists currently. We hope this paper will motivate others in the field to work through the details of such models. 

\indent In this paper, we constrain a broad class of ``hairy" BH models using a generic and model-independent EM signal that is characterized by the BH mass ($M$) and the fraction of that mass that is lost to EM radiation ($\epsilon$). In Section \ref{sec:schwinger}, we discuss the two phenomenologically distinct cases in which radiation is emitted and derive the critical value of $\epsilon$ that separates them. This division is set by the Schwinger limit, above which the BH radiation triggers pair production resulting in a GRB and below which the EM field accelerates ambient charged particles to create an overdensity of cosmic rays. In Section \ref{sec:gamma} we characterize the extragalactic observability of a GRB created by the BH radiation to constrain $\epsilon$ given the non-detection of GRBs from BBH mergers. In Section \ref{sec:GCRsignal} we describe the electron and proton cosmic ray energy spectrum created below the Schwinger limit and discuss the difficulties of observationally constraining $\epsilon$ in this less-energetic regime. In Section \ref{sec:conc} we summarize our results. 


\section{The Schwinger Limit} \label{sec:schwinger}

\indent The Schwinger limit dictates the critical value of $\epsilon$ separating the two phenomenologically distinct cases in which radiation is emitted for a particular BH mass. This limit, derived from quantum electrodynamics, sets the field strength at which an electric field becomes nonlinear due to the spontaneous production of electron-positron pairs \cite{Heisenberg_1936, Schwinger_1951}. Quantitatively, the Schwinger limit occurs at an electric field strength of $\mathcal{E}_C=m_e^2/e=0.86 \text{ MeV} ^{2}$. This corresponds to a field energy density of $u_C=\mathcal{E}_C^2=0.74 \text{ MeV} ^{4}$.

\indent This field energy density can be related to $\epsilon$ as follows. We assume the radiation from the BH is spread over a volume one wavelength ($\lambda$) in thickness outside of the BH Schwarzschild radius. Then, the energy density of the BH radiation is
\begin{eqnarray}
    u&&=\frac{\epsilon M}{\frac{4}{3}\pi ((r_s + \lambda)^3 - r_s^3)}\nonumber\\
    &&= 3.1\times10^9\epsilon \left(\frac{M}{M_\odot}\right)^{-2} \text{ MeV}^4.
\end{eqnarray}

Setting $u=u_C$ and solving for the critical value of $\epsilon$ gives
\begin{equation}
    \epsilon_C=2.4\times10^{-10}\left(\frac{M}{M_\odot}\right)^{2}.
\end{equation}
Therefore, $\epsilon_C\sim 10^{-10}-10^{-6}$ for BH masses ranging from 1 to 50 $M_\odot$. For $\epsilon>\epsilon_C$, pair-production dominates and results in a GRB as discussed in Section \ref{sec:gamma}. For $\epsilon<\epsilon_C$, the EM field accelerates ambient charged particles, creating cosmic ray electrons and protons as discussed in Section \ref{sec:GCRsignal}.


\section{Gamma-Ray Emission Above the Schwinger Limit} \label{sec:gamma}

\indent Above the Schwinger limit, the energy density of the field is large enough to result in electron-positron pair production \cite{Lieu_1998}. Qualitatively, the electron-positron-photon gas thermalizes due to Thompson scattering and expands relativistically as an ideal fluid, creating an object known as a fireball in GRB literature \cite{Goodman_1986,Paczynski_1986}. We denote the lab frame of an Earth observer as S and the comoving frame of the fluid as S'.

\indent When the EM radiation is first emitted by the BH, the lab and fluid frames coincide. The initial temperature of the fireball in both S and S' is given by 
\begin{equation}
    T_0=\left(\frac{E}{V_0 g_0 a}\right)^{1/4}
\end{equation}
where $a=\pi^2/15$, $E$ is the energy dumped into a region of volume $V_0$, and $g_0 = 2.75=11/4$ is half of the effective degrees of freedom for a plasma consisting of photons, electrons, and positrons in thermal equilibrium \cite{Goodman_1986}. Again assuming that the radiation from the BH is spread over a volume one wavelength in thickness outside of the BH Schwarzschild radius, the initial temperature is 
\begin{eqnarray}
    T_0&&=\left(\frac{\epsilon M}{\frac{4}{3}\pi((r_s+\lambda)^3-r_s^3)\frac{11}{4}a}\right)^{1/4}\nonumber\\
    &&=200\epsilon^{1/4}\left(\frac{M}{M_\odot}\right)^{-1/2} \text{ MeV}.
\end{eqnarray}

\indent Assuming that any remnants of stellar ejecta and envelopes have long dispersed, we neglect any external baryon contributions to the dynamics of the fireball. Thus, the fireball is a relativistic radiation-dominated fluid, which rapidly accelerates to $\gamma\gg1$ under its own super-Eddington radiation pressure. Because the fireball is created outside the Schwarzschild radius of the BH, the system originates in a region of small curvature. Consequently, general relativistic and gravitational redshift effects can be neglected. Employing the usual relativistic conservation equations of baryon number and energy-momentum from Ref. \cite{Weinberg_1972} in the limit where $\gamma\gg1$, yields the following scaling relations for each fluid shell \cite{Piran_1993,Kumar_2015}
\begin{align}
    \gamma(r)&\sim \left(\frac{r}{R_0}\right)\\
    T'(r)&\sim T_0\left(\frac{r}{R_0}\right)^{-1}\sim\frac{T_0}{\gamma(r)}
\end{align}
where $r$ is the distance from the origin in the lab frame and $R_0$ is the initial width of the fireball. These relations apply so long as the fireball is ultra-relativistic, radiation-dominated, and opaque due to Thompson scattering. So, as the fireball expands from the origin, the bulk Lorentz factor continues to increase due to the acceleration from the radiation pressure of the fluid. To first order in $\gamma$, the width of the fireball in the lab frame is constant $R(r)= R_0$ \cite{Meszaros_2006}. This requires the width in the comoving frame to increase as $R'(r)= \gamma(r)R_0$, illustrating why the fireball cools in its co-moving frame. In the lab frame, the temperature is blue-shifted by
\begin{equation}
    T(r)=\gamma(r)T'(r)= T_0
\end{equation}
since the fluid is moving relativistically towards the observer. Thus, so long as the scaling relations apply, a lab observer sees each shell of the fireball at the same constant temperature, $T_0$. 

\indent Within each fluid shell, the number density of electron-positron pairs in the comoving frame decreases as the fireball cools. Eventually, the process of pair creation and annihilation freezes out when the time for a positron to annihilate with an electron is of the same order as the dynamical time. This occurs at a comoving temperature of $T'\sim 20 \text{ keV}$ \cite{Piran_1999, Kumar_2015}. At this temperature, the proportion of the initial energy from the BH contained in the remaining electron-positron pairs is negligible \cite{Paczynski_1986}. So, nearly all of the initial $\epsilon M$ is contained in photons that had been trapped in the fluid by Thompson scattering. When pair production freezes out, the Thompson opacity decreases dramatically and these photons escape. Since the comoving temperature depends only on $r$, each shell experiences this freeze out as it moves through the same radius in the lab frame. Thus, the characteristic time delay between when photons free-stream from the inner and outermost edges of the fireball is given by \cite{Goodman_1986}
\begin{equation}
    \delta t\sim\frac{R_0}{c}=2.0\times10^{-5}\left(\frac{M}{M_\odot}\right)\text{ s}
\end{equation}
since the initial radius of the fireball is set by the wavelength of the BH radiation, $\lambda$. These timescales are short enough to be consistent with a short GRB, $<O(1$ second).

\begin{figure}[h!]
\begin{center}
\includegraphics[scale=0.37]{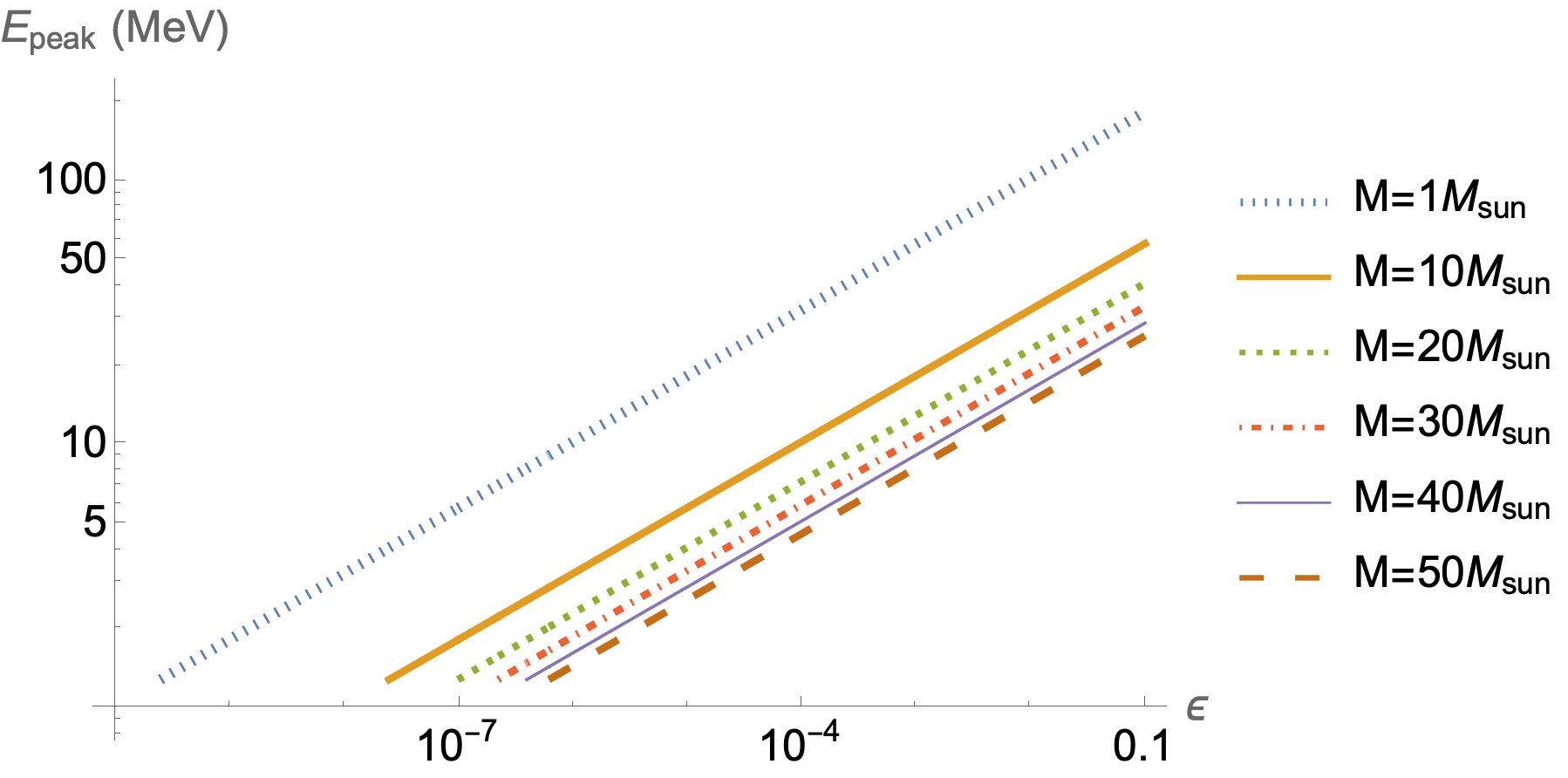}
\caption{The peak photon energy, $E_{peak}$, as a function of BH mass and $\epsilon$ for $\epsilon>\epsilon_C$. These peak energies are $>1$ MeV and so are gamma rays. \label{fig:Epeak}}
\end{center}
\end{figure}

\indent When the photons are able to free stream from the fireball, an observer on Earth sees a nearly thermal black body spectrum at temperature $T_0$ radiating from the BH \cite{Piran_1999}. The black body spectrum in photon number as a function of photon frequency and the fireball temperature is given by   
\begin{equation}
    B_f(T_0)=\frac{2f^2}{e^{2\pi f/T_0}-1}\text{ MeV}^2\text{ s}^{-1}\text{ Hz}^{-1}\text{ sr}^{-1}.
\end{equation}
The peak photon energy for this spectrum is
\begin{equation}
    E_{peak}=1.6 T_0=320\epsilon^{1/4}\left(\frac{M}{M_\odot}\right)^{-1/2} \text{ MeV}.
\end{equation}
These peak energies are plotted in Figure \ref{fig:Epeak} as a function of BH mass and $\epsilon$. These are the most likely photons to be emitted from the fireball, and all peak energies are gamma rays (energy $>1$ MeV). 

\indent Since nearly all the energy from the fireball is converted into photons at the peak energy, the number of photons emitted is 
\begin{equation}
    N_\gamma=\frac{\epsilon M}{E_{peak}}= 3.4\times10^{57}\epsilon^{3/4}\left(\frac{M}{M_\odot}\right)^{3/2}.
\end{equation}

\begin{figure}[h!]
\begin{center}
\includegraphics[scale=0.37]{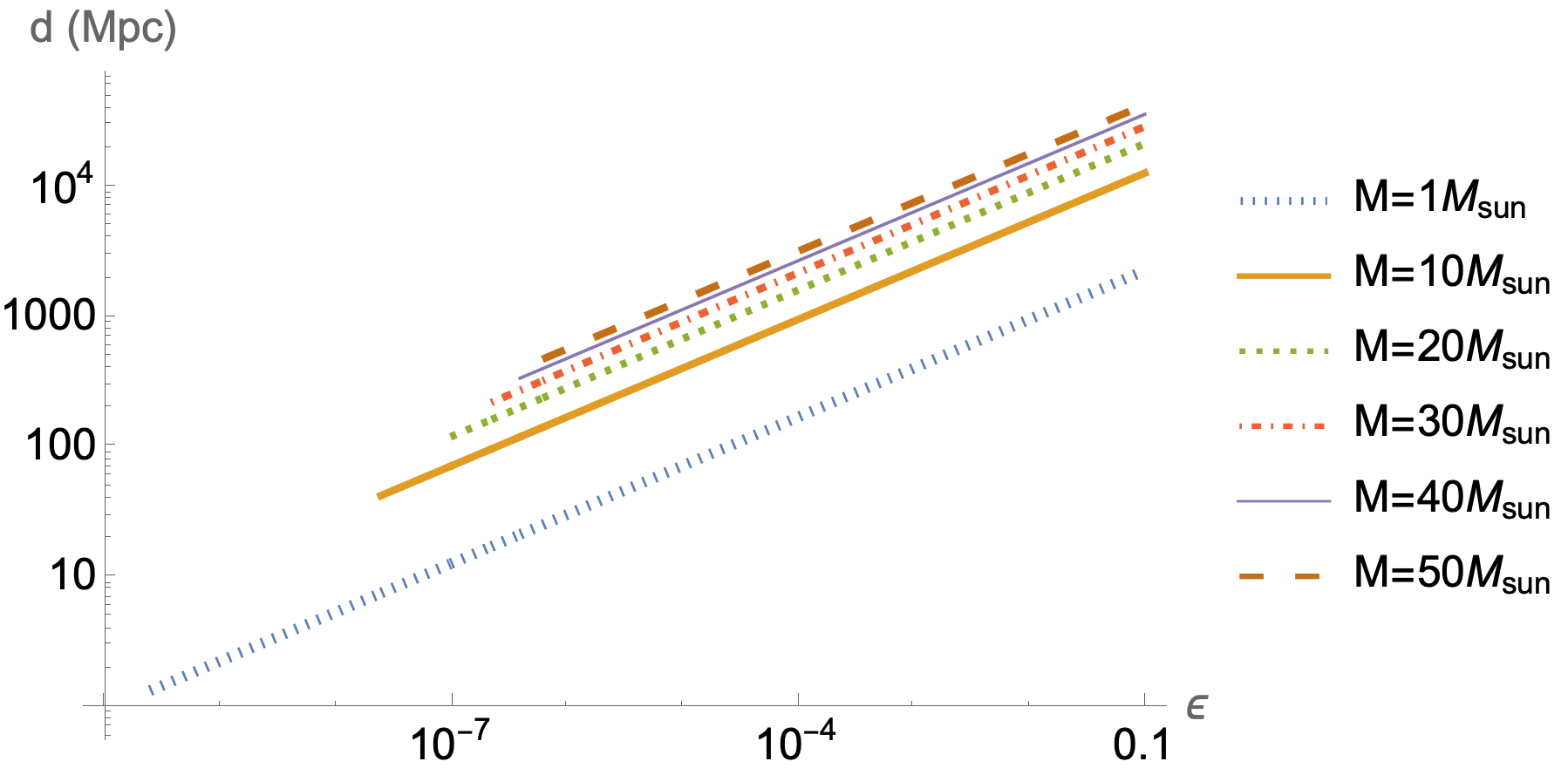}
\caption{The distance to which the gamma-ray photons emitted by 1-50 $M_\odot$ BHs are observable with Fermi, $d$, for various $\epsilon>\epsilon_C$. \label{fig:Distance}}
\end{center}
\end{figure}

\indent Over the full range of photon energies shown in Figure \ref{fig:Epeak}, the most sensitive telescope is the Fermi Gamma-ray Space Telescope. The two instruments onboard Fermi are the Large Area Telescope (LAT) and the Gamma-ray Burst Monitor (GBM). The LAT observes photon energies in the range $20$ MeV$-300$ GeV with a sensitivity of $10^{-4}$ erg cm$^{-2}$ \cite{Dermer_2013}. The GBM observes photon energies in the range $8$ keV$-40$ MeV with a sensitivity of $0.5$ ph cm$^{-2}$ s$^{-1}$ \cite{FermiGBM}. For the gamma-ray energies emitted by 1-50 $M_\odot$ BHs over a timescale of $\lesssim 1$ second, Fermi's sensitivity limits requires a flux of $F\gtrsim 1$ ph cm$^{-2}$. This minimum flux can be used to calculate the maximum distance, $d$, to which EM emission by a BH above the Schwinger limit is observable.
\begin{equation}
    d=\sqrt{\frac{N_\gamma}{4\pi F}}= 5360\epsilon^{3/8} \left(\frac{M}{M_\odot}\right)^{3/4}\text{ Mpc}
\end{equation}
These distances are plotted in Figure \ref{fig:Distance}. For a given distance, we can solve for the minimum $\epsilon$ for which Fermi could observe such an event extragalactically. 
\begin{equation}
    \epsilon_{min}=2\times10^{-5}\left(\frac{d}{100\text{  Mpc}}\right)^{8/3}\left(\frac{M}{M_\odot}\right)^{-2}
\end{equation}

\indent The Gravitational-wave (GW) Transient Catalog lists events detected by LIGO \cite{LIGO_2015}, Virgo \cite{Acernese_2015}, and the Kamioka Gravitational Wave Detector (KAGRA, \cite{Aso_2013}). Most BBH merger candidates listed in the catalog are posted on NASA's General Coordinates Network, which contains both concurrent and follow-up EM observations of GW triggers by the Fermi GBM. The Fermi GBM has an 8 steradian field-of-view \cite{FermiGBM}, which is large enough to cover the full LIGO/Virgo 90\% confidence GW localization region if the telescope is well-aligned at the time of the trigger.  We cross-reference all BBH mergers in the GW Transient Catalog with Fermi observations from the General Coordinates Network. For BH masses ranging from 10 to 50 $M_\odot$ in intervals of 10 $M_\odot$, we identify the nearest BBH merger for which Fermi observed at least 90\% of the localization region and recorded no GRB event. The GW events constraining each mass interval are listed in Table \ref{tab1}. These non-detections are used to constrain $\epsilon$ for each BH mass. To do this, we assume the furthest distance (luminosity distance + error bar) measured by LIGO/Virgo and calculate the minimum value of $\epsilon$ needed such that the event would be observable with the Fermi GBM for the observed BH mass. All constrained values of $\epsilon$ are above the Schwinger limit for the given BH mass and all result in $\sim$ MeV photons which are in the energy range observable by the GBM. These constraints are listed in Table \ref{tab2}. Assuming all BHs are ``hairy" BHs capable of producing this signal, the current upper bounds on $\epsilon$ are $\epsilon<10^{-5}$ for 10, 30, 40 $M_\odot$ BHs and $\epsilon<10^{-4}$ for 20, 50 $M_\odot$ BHs since no high energy EM signal was observed from these BBH mergers. These constraints will improve as more GW events with concurrent Fermi observations are detected.

{\renewcommand{\arraystretch}{1.5} 
\begin{table*}
\caption{\label{tab1}The table lists the physical parameters for the GW events constraining $\epsilon$ for each BH mass interval. GW191216\_213338-v1 was reported in Ref. \cite{GWTC3}, and all other events were reported in Ref. \cite{GWTC2}. The table provides the event ID from the GW Transient Catalog, the measured constituent BH masses ($m_1,$ $m_2$), the measured luminosity distance ($d_L$), the General Coordinates Network circular, and the percent of the localization region observed by the Fermi GBM (\% LR).}
\begin{ruledtabular}
\begin{tabular}{cccccc}
\textrm{ID}  & \textrm{$m_1/M_\odot$} & \textrm{$m_2/M_\odot$} & \textrm{$d_L$/Mpc} & \textrm{GCN Reference} & \textrm{\% LR} \\  \hline
GW191216\_213338 & $12.1_{-2.3}^{+4.6}$  & $7.7_{-1.9}^{+1.6}$  & $340_{-130}^{+120}$ & \cite{GCN_26454} & 99.8 \\ 
GW190915\_235702 & $32.6_{-4.9}^{+8.8}$  & $24.5_{-5.8}^{+4.9}$ & $1750_{-650}^{+710}$ & \cite{GCN_25752}  & 98.0 \\ 
GW190412\_053044 & $27.7_{-6.0}^{+6.0}$  & $9.0_{-1.4}^{+2.0}$    & $720_{-220}^{+240}$  & \cite{GCN_24098} & 99.8 \\ 
GW190521\_074359 & $43.4_{-5.5}^{+5.8}$  & $33.4_{-6.8}^{+5.2}$ & $1080_{-530}^{+580}$ & \cite{GCN_24629} & 100.0 \\ 
GW190701\_203306 & $54.1_{-8.0}^{+12.6}$ & $40.5_{-12.1}^{8.7}$ & $2090_{-740}^{+770}$ & \cite{GCN_24948} & 100.0 \\ 
\end{tabular}
\end{ruledtabular}
\end{table*}}

{\renewcommand{\arraystretch}{1.5} 
\begin{table*}
\caption{\label{tab2} This table lists the constraints on $\epsilon$, assuming all BHs are ``hairy" BHs, for each mass range. The columns give the event ID from the GW Transient Catalog, the BH mass range ($O(m)$), the maximum luminosity distance ($d_{L, Max}$), the minimum value of $\epsilon$ such that the event would be observable ($\epsilon_{Min}$), and the peak photon energy ($E_{peak}$) for that $\epsilon_{min}$ and BH mass.}
\begin{ruledtabular}
\begin{tabular}{ccccc}
\textrm{ID} & \textrm{$O(m)/M_\odot$} & \textrm{$d_{L, Max}$/Mpc} & \textrm{$\epsilon_{Min}$}   & \textrm{$E_{peak}$/MeV} \\ \hline
GW191216\_213338 & 10 & 460 & $1.4\times10^{-5}$ & 6.3             \\ 
GW190915\_235702 & 20 & 2460 & $3.1\times10^{-4}$ & 9.6             \\ 
GW190412\_053044 & 30 & 960 & $1.1\times10^{-5}$ & 3.4             \\ 
GW190521\_074359 & 40 & 1660 & $2.7\times10^{-5}$ & 3.7             \\ 
GW190701\_203306 & 50 & 2860 & $7.5\times10^{-5}$ & 4.3             \\ 
\end{tabular}
\end{ruledtabular}
\end{table*}}

\indent There has been one observation of a GRB and BBH merger occurring concurrently. An offline search of Fermi GBM data following the detection of GW150914 \cite{Abbott_2016} revealed a 1 second short GRB occurring 0.4 seconds after the LIGO trigger with a localization consistent with that of the GW signal \cite{Connaughton_2016}.  The GRB transient was near the detection threshold for the GBM (2.9$\sigma$ detection), was not detected by any other instrument, and the localization was poorly constrained \cite{Connaughton_2016}. Thus, the GRB cannot be confidently associated with GW150914. Assuming that the GRB did indeed originate from the BBH merger, we can identify the range of $\epsilon$ that is consistent with the measured properties of the merger and GRB. From the GW Transient Catalog, the constituent BH masses range from $\sim$30-40 $M_\odot$ and the range of luminosity distances is 270-590 Mpc. For these masses and distances, the minimum value of $\epsilon$ such that the event is observable with the Fermi GBM ranges from $\epsilon_{min}\sim10^{-7}-10^{-6}$, resulting in peak photon energies of $\sim 1-3$ MeV. This range of peak photon energies is consistent with the properties of the observed GRB, which peaked near an MeV \cite{Connaughton_2016}. Since this event was barely above the GBM's detection threshold, we anticipate $\epsilon\sim\epsilon_{min}$. Therefore, the observed GRB is consistent with a GRB produced via rapid EM emission directly from a 30-40 $M_\odot$ ``hairy" BH for $\epsilon\sim10^{-7}-10^{-6}$.


\section{Galactic Cosmic Ray Signal Below the Schwinger Limit} \label{sec:GCRsignal}

\indent Below the Schwinger limit the electric field emitted by the BH propagates outwards, accelerating ambient charged particles. Mutual attraction between the protons and electrons inhibits charge separation, requiring that the particles have the same Lorentz factor on average. Since the protons are a factor of $\sim10^3$ heavier than the electrons and the particle energy scales with mass, most of the BH field energy is absorbed by protons. Thus, the dynamics of the system are set by the protons.

\indent A rapid burst of isotropic EM radiation can be generically described by a coherent single-wavelength pulse. The exact duration and coherence of emission is model-dependent, but so long as the emission occurs on the short timescales characteristic of BBH mergers, the signal is sufficiently similar to the single-wavelength pulse approximation. First, we consider the acceleration of a single charged particle due to this strong EM pulse. In the (+ - - -) metric, the relativistic Lorentz force law is 
\begin{equation}
    \frac{du^\alpha}{d\tau}=\frac{e}{m}F^{\alpha\beta}u_\beta
\end{equation}
where $e$ is the elementary charge, $m$ is the proton or electron mass, $u^\alpha$ is the particle's 4-velocity, and $\tau$ is the proper time in the instantaneous rest frame of the particle. For a sinusoidal EM field, the characteristic timescale of field variation is $\mathcal{E}/\frac{d\mathcal{E}}{dt}\sim 1/f$ where $\mathcal{E}$ is the electric field magnitude and $f$ is the frequency of the BH radiation. Since $f\sim10^{-17}$ MeV is very small, this characteristic timescale is very long. So, the field magnitude is nearly constant in time and the system can be approximated by a constant crossed field. For the ranges of $\epsilon$ considered here, the absorption length for the EM field is long enough that the average charged particle experiences a sufficiently weak field to neglect any special relativistic effects that transform the field but also a sufficiently strong field that the protons are quickly accelerated to $v\sim 1$ and radiate negligibly. We choose a coordinate system such that the electric field is directed along the x-axis and the magnetic field along the y-axis, then the Faraday tensor is 
\begin{equation}
    F^{\alpha\beta}=\begin{pmatrix}
0 & -\mathcal{E} & 0 & 0\\
\mathcal{E} & 0 & 0 & 0\\
0 & 0 & 0 & \mathcal{E}\\
0 & 0 & -\mathcal{E} & 0
\end{pmatrix}
\end{equation}
where $\mathcal{E}$ is the magnitude of the electric field in MeV$^2$. Since $\frac{e}{m}F^{\alpha\beta}$ is constant, the Lorentz force equation has a matrix exponential solution
\begin{equation}
    u^\alpha(\tau)=\exp{\left(\frac{e}{m}\tau F^{\alpha}_{\text{ }\beta}\right)}u^\beta(0).
\end{equation}
The particles accelerate from the thermal speed of the plasma ($v\sim 0$), fixing $u^\beta(0)$.
We compute $u^\alpha(\tau)$ for the given Faraday tensor and use this to extract the Lorentz factor, $\gamma$, and the components of the 3-velocity in the lab frame, $\textbf{v}$.
\begin{align}
    \gamma(\tau)&=1+\frac{e^2\mathcal{E}^2\tau^2}{2m^2}\\
    v_x(\tau)&=\frac{2e\mathcal{E}m\tau}{2m^2+e^2\mathcal{E}^2\tau^2}\\
    v_y(\tau)&=0\\
    v_z(\tau)&=1-\frac{2m^2}{2m^2+e^2\mathcal{E}^2\tau^2}
\end{align}

\indent The final kinetic energy of the particle depends on the time at which the particle exits the field in its instantaneous rest frame, $\tau_f$. By definition, $\gamma(\tau)=\frac{dt}{d\tau}$. Thus, 
\begin{equation}
    \int_0^{\tau_f}\gamma(\tau)d\tau=\tau_f+\frac{e^2\mathcal{E}^2\tau_f^3}{6m^2}=\int_0^{t_f}dt=\frac{1}{f}
\end{equation}
since in the lab frame the particle is in the field for $t_f=\frac{1}{f}$. The magnitude of the EM field is sufficiently large, and we can neglect the term that is linear in $\tau_f$ to find
\begin{eqnarray}
    \tau_f&&=\left(\frac{6m^2}{e^2\mathcal{E}^2f}\right)^{1/3}\nonumber\\
    &&= 1.2\times10^8\frac{1}{\mathcal{E}^{2/3}}\left(\frac{M}{M_\odot}\right)^{1/3} \text{ MeV}^{-1},
\end{eqnarray}
where here, and in all following numerical expressions, we set $m=m_p$.

\indent Now we characterize the absorption of energy from the BH EM field by ambient protons. We partition the volume around the BH into spherical shells one wavelength in thickness such that the distance from the BH is parameterized by the dimensionless number $j$, the number of wavelengths from the Schwarzschild radius of BH. Let $E$ be the total EM energy incident onto a shell $j$ wavelengths from the BH. Assuming $j$ is large, the initial energy density of the shell is

\begin{eqnarray}
    u&&= \frac{E}{\frac{4}{3}\pi[(r_s+j\lambda)^3-(r_s+(j-1)\lambda)^3]}\nonumber\\
    &&\sim \frac{E}{4 \pi \lambda^3 j^2}\nonumber\\
    &&= 3.0\times10^{-51}\frac{E}{ j^2}\left(\frac{M}{M_\odot}\right)^{-3}\text{ MeV}^4.
\end{eqnarray}
Then the magnitude of the electric field is
\begin{eqnarray}
    \mathcal{E}&&=\sqrt{u}\nonumber\\
    &&=5.4\times10^{-26} \left(\frac{E}{j^2}\right)^{1/2}\left(\frac{M}{M_\odot}\right)^{-3/2}\text{ MeV}^2.
\end{eqnarray}

\indent The kinetic energy of each proton after interacting with the EM wave is
\begin{eqnarray}
    K&&\sim\gamma(\tau_f)m\nonumber\\
    &&= 1.0\times 10^{-5}\left(\frac{E}{j^2}\right)^{1/3}\left(\frac{M}{M_\odot}\right)^{-1/3} \text{ MeV}.
\end{eqnarray}

This gives the kinetic energy of one proton in the $j$th shell. To derive the total kinetic energy lost in the $j$th shell, we assume a homogeneously distributed number density of charged particles with $n=n_p=n_e\sim 1\text{ cm}^{-3}\sim 10^{-32}\text{ MeV}^{3}$, consistent with the Milky Way interstellar medium \cite{Canto_1977, Reynolds_1992}. The total number of protons in the $j$th shell is
\begin{equation}
    N\sim 4 \pi n \lambda^3 j^2=2.6\times10^{18}j^2 \left(\frac{M}{M_\odot}\right)^3.
\end{equation}
So the total kinetic energy absorbed in the $j$th shell is
\begin{equation}
    K_{tot}=N K=2.6\times10^{13}j^{4/3}E^{1/3}\left(\frac{M}{M_\odot}\right)^{8/3} \text{ MeV}.
\end{equation}

\indent This gives a differential equation for the energy absorbed by the field
\begin{equation}
    \frac{d E}{d j}=-K_{tot}
\end{equation}
subject to the initial condition $E(0)=\epsilon M$. This equation can be integrated to yield

\begin{widetext}
\begin{equation}
    E(j)=\left(1.1\times10^{40}\epsilon^{2/3}\left(\frac{M}{M_\odot}\right)^{2/3}-7.5\times10^{12}\left(\frac{M}{M_\odot}\right)^{8/3}j^{7/3}\right)^{3/2}\text{ MeV}.
\end{equation}
\end{widetext}

Solving for the value of $j$ at which $E(j)=0$ gives the dimensionless absorption length, $j_{abs}$,
\begin{equation}
    j_{abs}= 4.4\times10^{11}\epsilon^{2/7}\left(\frac{M}{M_\odot}\right)^{-6/7}.
\end{equation}

The absorption length, $j_{abs}\lambda$, gives the number of protons accelerated by the field and the average kinetic energy and Lorentz factor of each proton.
\begin{align}
    N_{abs}&=\frac{4}{3}\pi n (j_{abs}\lambda)^3=7.1\times10^{52}\epsilon^{6/7}\left(\frac{M}{M_\odot}\right)^{3/7}\\
    K_{avg}&=\frac{\epsilon M}{N_{abs}}=1.6\times10^7\epsilon^{1/7}\left(\frac{M}{M_\odot}\right)^{4/7}\text{ MeV}\\
    \gamma_{avg}&\sim\frac{KE_{avg}}{m}=1.7\times10^4\epsilon^{1/7}\left(\frac{M}{M_\odot}\right)^{4/7}
\end{align}
The number of electrons accelerated by the field is also $N_{abs}$ and the average Lorentz factor of the electrons must be the same as that for the protons. 

\begin{figure}[h!]
\begin{center}
\includegraphics[scale=0.37]{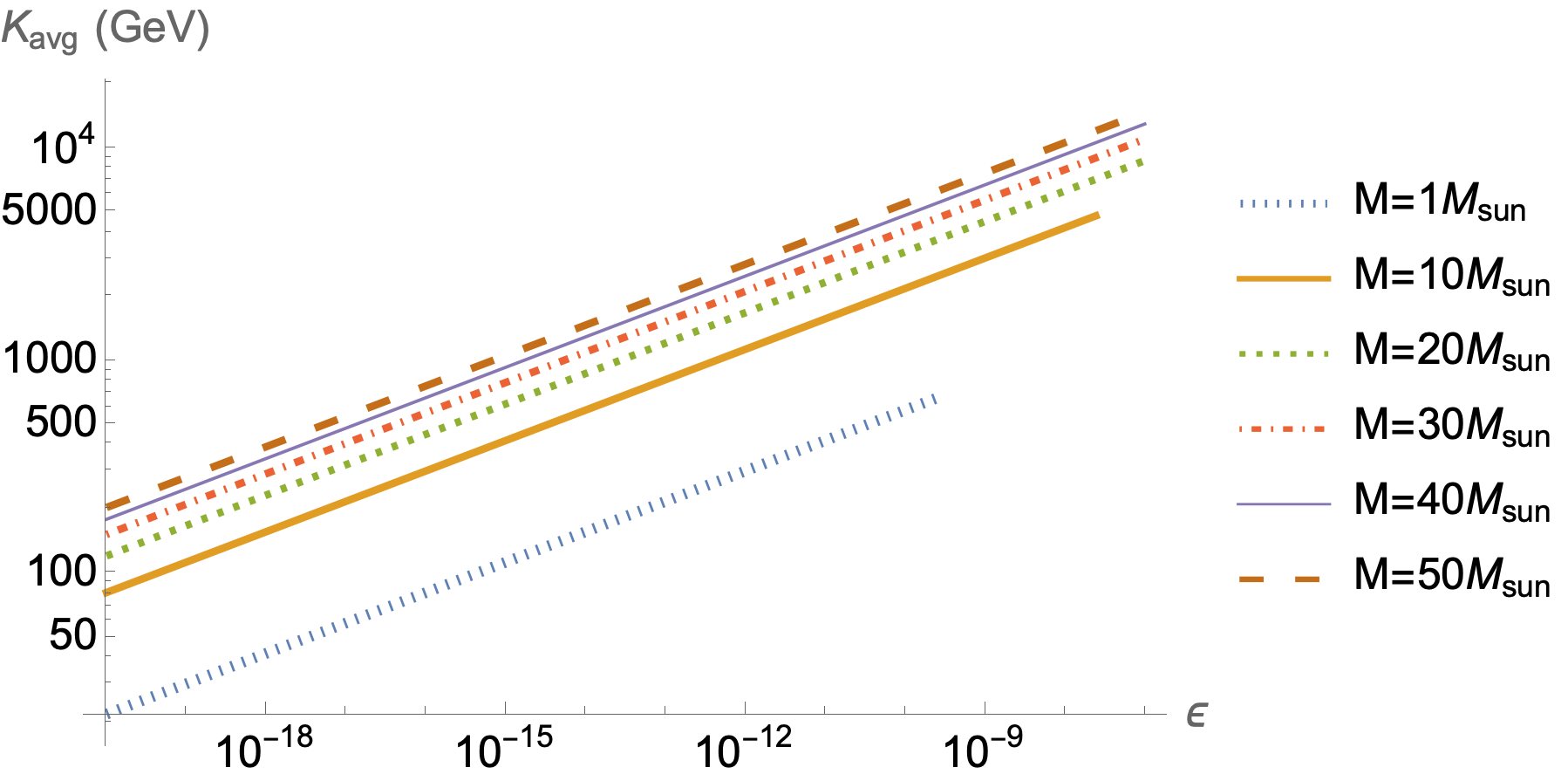}
\caption{The average proton kinetic energy, $K_{avg}$, for 1-50 $M_\odot$ BHs as a function of $\epsilon$ for $\epsilon<\epsilon_C$. These relativistic protons are at cosmic ray energies. \label{fig:KEavg}}
\end{center}
\end{figure}

\indent We perform this calculation for values of $\epsilon$ ranging from $\epsilon=10^{-20}$ to the Schwinger limit for 1-50 $M_\odot$ BHs. The resulting average proton kinetic energies as a function of $\epsilon$ are plotted in Figure \ref{fig:KEavg}. Overall, the average kinetic energy per proton ranges from 20 GeV for a 1 $M_\odot$ BH with $\epsilon=10^{-20}$ to 20 TeV for a 50 $M_\odot$ BH at the Schwinger limit ($\epsilon\sim10^{-6}$). The average proton kinetic energies can be used to calculate the corresponding average electron kinetic energies. Overall, the average kinetic energy per electron ranges from 0.01 GeV for a 1 $M_\odot$ BH with $\epsilon=10^{-20}$ to 10 GeV for a 50 $M_\odot$ BH at the Schwinger limit. These relativistic protons and electrons are cosmic rays.

\indent Cosmic rays of these energies are difficult to identify with a point source on the sky. Ambient magnetic fields confine these cosmic rays to their host galaxies and cause them to quickly diffuse (on timescales $\lesssim 0.01$ years) into the galactic background of cosmic rays. Thus, these cosmic rays are indistinguishable from cosmic rays produced by supernova remnants \cite{Dermer_2013}, stellar winds or flares \cite{Miroshnichenko_2001}, and other processes. Although these particles create secondary signals as they lose energy to bremsshtrahlung, ionization, synchrotron radiation and inverse Compton scattering for electrons and to inelastic collisions for protons \cite{Longair_1992, Longair_1994}, the timescales for these processes is much longer than the diffusion time ($\gtrsim 10^6$ years). Therefore, these secondary signals are also indistinguishable from those produced by other cosmic ray processes. 

\indent Because these cosmic rays are mixed with and are indistinguishable from cosmic rays produced via other mechanisms, it is not possible to confidently identify cosmic rays due to EM radiation from BHs either in external galaxies or in the Milky Way. Therefore, it becomes difficult to place strong constraints on $\epsilon$ below the Schwinger limit. Although this avenue cannot constrain $\epsilon$, the effect is still potentially observable. Should the BBH merger occur on a sufficiently strong magnetic field background ($B>10$ MeV$^2=0.05$ T), the ultrarelativistic electrons would produce synchrotron radiation in the X-ray band, motivating X-ray observations of BBH mergers.


\section{Discussion and Conclusions} \label{sec:conc}

\indent The theoretical understanding of BHs is inconsistent, motivating searches for generic signals of deviations from canonical BH models. In this paper, we have constrained a broad class of ``hairy" BH models capable of emitting a fraction of their mass as EM radiation. Since this radiation is sourced directly from the BH, it must tunnel out of the BH's gravitational well in the same manner as Hawking radiation. Thus, the characteristic frequency of the radiation depends only on the mass of the BH, resulting in a signal that is generic and model-independent. We derive the critical value of $\epsilon$, the fraction of the BH mass released as radiation, above which the field strength triggers a GRB and below which ambient particles are accelerated to cosmic ray energies. Because no extragalactically-observable EM signal is expected from a stellar-mass BBH merger, we find that concurrent observations of BBH mergers with GW detectors and EM radiation instruments offer the best data to detect such a signal. 

\indent In the GRB regime, the BH mass and $\epsilon$ fix the initial volume and temperature of the electron-positron fireball. The fireball expands relativistically, maintaining constant temperature in the frame of an Earth observer and cooling in its comoving frame. Once the fireball is sufficiently cool in its frame, pair-production freezes out and the photons free stream. In the frame of an Earth observer, these photons have energies described by a black body spectrum at the initial temperature of the fireball. Thus, the energy deposited by the BH is re-emitted as gamma-rays over a short timescale. By cross-referencing GW events with concurrent Fermi GBM observations of the localization region, we place upper bounds on $\epsilon$. These bounds are $\epsilon<10^{-5}-10^{-4}$ for $10-50$ $M_\odot$ BHs depending on the BH mass since no high energy EM signal was observed from these BBH mergers. These constraints will improve as more GW events with concurrent Fermi observations are detected. We also discuss the weak detection of a GRB following GW150914, and find that this event is consistent with a GRB produced via rapid EM emission directly from a ``hairy" BH for $\epsilon\sim 10^{-7}-10^{-6}$. 

\indent Below the Schwinger limit, the EM radiation can be described by a constant-crossed field. The dynamics of the system are fixed by the ambient protons, which are rapidly accelerated to $v\sim 1$ by the field, absorbing energy as the radiation propagates away from the BH. We solve the differential equation describing the energy lost by the field to calculate the absorption length as a function of BH mass and $\epsilon$. This absorption length fixes the average energy of the ambient protons and electrons that interacted with the BH radiation. For 1-50 $M_\odot$ BHs and $\epsilon$ ranging from $10^{-20}$ to the Schwinger limit, the average kinetic energy per proton ranges from 20 GeV-20 TeV and the energy per electron ranges from 0.01-10 GeV. At these energies, cosmic rays have a short diffusion length due to the galactic magnetic field and are mixed in with other astrophysical cosmic rays. Additionally, the secondary signals from cosmic rays of these energies are produced on too long of a timescale to be attributed to EM radiation directly from a BH. Overall, constraining $\epsilon$ in this less energetic regime is difficult. Future work could investigate BBH mergers in strong background magnetic fields. In this case, the ultrarelativistic electrons emit X-rays via synchrotron radiation that may be observable.

\indent Although this work benefits from employing a model-independent approach to generically characterize radiation emitted directly from ``hairy" BHs, model-dependent effects will need to be included to augment this general parameterization. Some ``hairy" BH models, such as the firewall BH \cite{Kaplan_2019}, are capable of producing EM radiation. But currently, no complete model able to quantitatively characterize this effect exists. We hope this paper will motivate others in the field to work through the details of such models. The firewall BH metric of Ref. \cite{Kaplan_2019} is particularly well-suited for this. Given this metric and a parameterized charge distribution adhered to the firewall, one could use numerical relativity to simulate the emission of gravitational and EM radiation during a BBH merger involving a firewall BH. This approach offers independent constraints on $\epsilon$ as a function of the charge distribution parameter and may be able to constrain $\epsilon$ in cases below the Schwinger limit.

\indent Strengthening constraints on $\epsilon$ will be an important pursuit for future work given the relatively loose bounds found in this paper. In general, ``hairy" BHs are well-motivated by the BH information paradox. However, such BHs must appear nearly canonical to a distant observer due to observational evidence in favor of the ``no hair" theorem. Therefore, novel approaches to constraining ``hairy" models will continue to be vital in the search for new fundamental physics.


\acknowledgments

\indent We thank Surjeet Rajendran and David Kaplan for useful discussions. We also thank Nadia Zakamska and Erwin Tanin for their edits to this manuscript. 

\indent This work is supported by the National Science Foundation Graduate Research Fellowship Program under Grant No. DGE2139757. Any opinions, findings, and conclusions or recommendations expressed in this material are those of the author and do not necessarily reflect the views of the National Science Foundation.


\end{document}